\documentclass[preprint,aip,amsbsy,amsmath,amssymb,amsart,amsintx]{revtex4-1}

\usepackage[latin1]{inputenc}
\usepackage[active]{srcltx}
\usepackage{setspace}
\usepackage{graphicx}
\usepackage{dcolumn}
\usepackage{bm}
\usepackage{epsfig}
\usepackage{epstopdf}

\begin{document}


\title[]{A Variational Method in Out of Equilibrium Physical Systems}

\author{Mario J. Pinheiro} \email{mpinheiro@ist.utl.pt}
\homepage{http://web.ist.utl.pt/ist12493}
\affiliation{Department of Physics, Instituto Superior T\'{e}cnico, Av. Rovisco Pais,
1049-001 Lisboa, Portugal \\
\& Institute for Advanced Studies in the Space, Propulsion and Energy Sciences \\
265 Ita Ann Ln.
Madison, AL 35757
USA
}

\homepage{http://www.ias-spes.org}

\thanks{}

\date{\today}

\begin{abstract}
A variational principle is further developed for out of equilibrium dynamical systems by using the concept of maximum entropy. With this new formulation it is obtained a set of two first-order differential equations, revealing the same formal symplectic structure shared by classical mechanics, fluid mechanics and thermodynamics. In particular, it is obtained an extended equation of motion for a rotating dynamical system, from where it emerges a kind of topological torsion current of the form $\epsilon_{ijk} A_j \omega_k$, with $A_j$ and $\omega_k$ denoting components of the vector potential (gravitational or$/$and electromagnetic) and $\omega$ is the angular velocity of the accelerated frame. In addition, it is derived a special form of Umov-Poynting's theorem for rotating gravito-electromagnetic systems, and obtained a general condition of equilibrium for a rotating plasma. The variational method is then applied to clarify the working mechanism of some particular devices, such as the Bennett pinch and vacuum arcs, to calculate the power extraction from an hurricane, and to discuss the effect of transport angular momentum on the radiactive heating of planetary atmospheres. This development is seen to be advantageous and opens options for systematic improvements.
\end{abstract}



\maketitle

\section{Introduction}

In the years 1893-96, the Norwegian explorer Fridtjof Nansen, while he were traveling in the Arctic region, noticed the drift of surface ice across the polar sea making an angle of 20 to 40 degrees to the right related to the wind direction. Nansen has advanced an explanation, that in addition to the wind's force, it was also actuating the Coriolis force. In 1905, Vagn Walfrid Ekman~\cite{Ekman_05} introduced a theory of wind currents in open seas showing that sea current change of direction with depth due to Coriolis force appearing in a rotating based coordinate system. Besides attracting considerable interest in geophysical flow problems~\cite{Greenspan}, these discoveries stimulated further investigation in such fields as magnetic geodynamics, binary stars, new-born planetary systems, trying to unveil the important problem of angular momentum transport, and as well, this work.

The problem of enhanced angular moment transport in accretion disks~\cite{Balbus 2003}, and the breakdown of Keplerian rotation, as well as the removing of angular momentum from a vortex due to moving spiral waves, playing an important role on the total angular momentum balance of the core and the intensification of a tropical cyclone~\cite{Chow 2004}, are all examples of problems that demand a clear understanding of the dynamics of electromagnetic-gravitational rotating systems. Furthermore, special attention must be dedicated to the role of flux of angular momentum, and its conservation. Notwithstanding all attempts to take an accurate account of the transport angular momentum~\cite{Kronberg}, it lacks an additional equation of conservation (besides continuity, momentum and energy equations), which is related to local density of angular momentum, and its related flux. In order to develop a logically closed foundation, the equation for the angular momentum balance must be included and, to our best knowledge, these problems were treated by Curtiss~\cite{Curtiss 1956} and Livingston and Curtiss~\cite{Livingston 1959}, and they are also addressed in the present work but using a variational method.

In this article, we develop a standard technique for treating a physical system on the basis of an information-theoretic framework previously developed~\cite{Jaynes,Pinheiro:02,Pinheiro:04}, which constitutes an entropy maximizing method leading to a set of two first-order differential equations revealing the same formal symplectic structure shared by classical mechanics and thermodynamics~\cite{Peterson_1979,TMP_Sergeev_2008}. Our method bears some resemblance with isoentropic but energy-nonconserving variational principle proposed by R. Jackiw {\it et al.}~\cite{Jackiw_1988} allowing to study nonequilibrium evolution in the context of quantum field theory, but with various classical analogous such as the Schr\"{o}dinger equation giving rise to reflectionless transmission.

The plan of the article is as follows. In Sec. II, we include the extended mathematical formalism to nonequilibrium information theory. In Sec. III, we analysis the equilibrium and stability of a rotating plasma. In Sec. IV, we apply our formalism to angular momentum transport, obtaining in particular the Umov-Poynting's theorem for rotating electromagnetic-gravitational systems (e.g., rotating plasmas, magnetic geodynamics, vortex motion and accretion disks in astrophysics), whose applications might contribute to clarify still poorly understood phenomena.

\section{Mathematical procedure}

Let us consider here a simple dynamical system constituted by a set of $N$
discrete interacting mass-points $m^{(\alpha)}$ ($\alpha=1,2,...,N$)
with $x_i^{\alpha}$ and $v_i^{\alpha}$ ($i=1,2,3; \alpha=1,...,N$)
denoting the coordinates and velocities of the mass point in a given
inertial frame of reference. The inferior Latin index refers to the
Cartesian components and the superior Greek index distinguishes the
different mass-points.

The gravitational potential $\phi^{(\alpha)}$ associated to the
mass-point $\alpha$ is given by
\begin{equation}\label{eq1}
\phi^{(\alpha)} = G \sum_{\substack{\beta=1\\ \beta \neq \alpha}}
\frac{m^{(\alpha)}}{|\mathbf{x}^{(\alpha)} - \mathbf{x}^{(\beta)}|},
\end{equation}
with $G$ denoting the gravitational constant and
$\mathbf{x}^{(\alpha)}$ and $\mathbf{x}^{(\beta)}$ representing
the instantaneous positions of the mass-points $(\alpha)$ and
$(\beta)$. $\sum$ denotes summation over all the particles of the system. Energy, momentum and angular momentum conservation laws must be verified for a totally isolated
system:
\begin{equation}\label{2a}
  E = \sum_{\alpha=1}^N E_{(\alpha)},
\end{equation}
\begin{equation}\label{2}
  \mathbf{P} = \sum_{\alpha=1}^N \mathbf{p}^{(\alpha)},
\end{equation}
the particles total angular momentum (sum of the orbital angular momentum plus the intrinsic angular momentum or spin)
\begin{equation}\label{3}
\mathbf{L} = \sum_{\alpha=1}^N \mathbf{L}^{(\alpha)} = \sum_{\alpha=1}^N ([\mathbf{r}^{(\alpha)} \times
\mathbf{p}^{(\alpha)}] + \mathbf{J}^{(\alpha)}).
\end{equation}
In the above equations, $\mathbf{r}^{(\alpha)}$ is the position vector
relatively to a fixed frame of reference $\mathcal{R}$,
$\mathbf{p}^{(\alpha)}$ is the total momentum (particle + field) and
$\mathbf{L}^{(\alpha)}$ is the total angular momentum of the
particle, comprising a vector sum of the particle orbital angular
momentum and the intrinsic angular momentum $\mathbf{J}$ (e.g., contributed by the electron spin and/or
nuclear spin, since the electromagnetic momentum is already
included in the preceding term through $\mathbf{p}^{(\alpha)}$). The maximum entropy principle introduces Lagrange
multipliers from which, as we will see, ponderomotive forces are
obtained.

It is necessary to find the conditional extremum; they are set up not for the function
$S$ itself but rather for the changed function $\bar{S}$. Following the mathematical procedure proposed in
Ref.~\cite{Pinheiro:04} the total entropy of the system
$\overline{S}$ is thus given by
\begin{eqnarray}\nonumber
\overline{S} &=& \sum_{\alpha=1}^N  \left\{ S^{(\alpha)} \left( E^{(\alpha)}
- \frac{(\mathbf{p}^{(\alpha)})^2}{2 m^{(\alpha)}}
- \frac{(\mathbf{J}^{(\alpha)})^2}{2 I^{(\alpha)}}
- q^{(\alpha)} V^{(\alpha)}
+ q^{(\alpha)} (\mathbf{A}^{(\alpha)} \cdot \mathbf{v}^{(\alpha)})
- U^{(\alpha)}_{mec} \right) \right. \\ \label{eq2}
&& \left. + (\mathbf{a} \cdot \mathbf{p}^{(\alpha)}
+ \mathbf{b} \cdot ([\mathbf{r}^{(\alpha)}
\times \mathbf{p}^{(\alpha)}] + \mathbf{J}^{(\alpha)}) \right\} = \sum_{\alpha=1}^N \mathfrak{S}^{(\alpha)}.
\end{eqnarray}
where $\mathbf{a}$ and $\mathbf{b}$ are (vectors) Lagrange multipliers. It can be shown that $\mathbf{v}_{rel}=\mathbf{a}T$ and $\pmb{\omega}=\mathbf{b}T$ (see also Ref.~\cite{Landau_1}). The conditional extremum points form the dynamical equations of motion of a general physical system (the equality holds whenever the physical system is in thermodynamic and mechanical equilibrium), defined by two first order differential equations:

\begin{subequations}\label{eq15d}
\begin{align}
\frac{\partial \bar{S}}{\partial \mathbf{p}^{(\alpha)}} \geq 0 & \blacktriangleright  ~\mbox{canonical momentum;} \label{g2b}\\
\nonumber \\
\frac{\partial \bar{S}}{\partial \mathbf{r}^{(\alpha)}} = -\frac{1}{T} \pmb{\nabla}_{r^{(\alpha)}} U^{(\alpha)} - \frac{1}{T} m^{(\alpha)} \frac{\partial \mathbf{v}^{(\alpha)}}{\partial t} \geq 0 & \blacktriangleright  ~\mbox{fundamental equation of dynamics}  \label{g2a}.
\end{align}
\end{subequations}

Eq.~\ref{g2a} gives the fundamental equation of dynamics and has the form of a general local balance equation having as source term the spatial gradient of entropy, $\nabla_a S>0$, whilst Eq.~\ref{g2b} gives the canonical momentum (see also Eq.~\ref{eq2ab}). At thermodynamic equilibrium the total entropy of the body has a maximum value, constrained through the supplementary conditions~\ref{2a}, ~\ref{2}, and ~\ref{3}, which ensues typically from the minimization techniques associated to the Lagrange multipliers. In the more general case of a non-equilibrium process, the entropic gradient must be positive in Eq.~\ref{g2a}, according to Vanderlinde's proposition ~\cite{Vanderlinde_2011}, a condition required for the gravitational force to exist.
However, new physics may be brought by the set of two first order differential equations, related to the interplay between the tendency of energy to attain a minimum, whilst entropy seeks to maximize its value.

In non-equilibrium processes the gradient of the total entropy in momentum space multiplied by factor $T$ is given by
\begin{equation}\label{6}
T \frac{\partial \overline{S}}{\partial \mathbf{p}^{(\alpha)}} = \left\{
-\frac{\mathbf{p}^{\alpha)}}{m^{(\alpha)}} + \frac{q^{(\alpha)}}{m^{(\alpha)}}\mathbf{A} +
\mathbf{v}_e + [\mathbf{\omega} \times \mathbf{r}^{(\alpha)}] \right\},
\end{equation}
so that maximizing entropy change in Eq.~\ref{g2b} leads to the well-known total (canonical)
momentum:
\begin{equation}\label{}
\mathbf{p}^{(\alpha)} = m^{(\alpha)} \mathbf{v}_e + m^{(\alpha)} [\mathbf{\omega} \times
\mathbf{r}^{(\alpha)}] + q^{(\alpha)} \mathbf{A}.
\end{equation}

The above formulation bears some resemblance with Hamiltonian formulation of dynamics which expresses first-order constraints of the Hamiltonian $H$ in a $2n$ dimensional phase space, $\dot{\mathbf{p}}=-\partial H/\partial \mathbf{q}$ and  $\dot{\mathbf{q}}=\partial H/\partial \mathbf{p}$, and can be solved along trajectories as quasistatic processes, revealing the same formal symplectic structure shared by classical mechanics and thermodynamics~\cite{Peterson_1979,TMP_Sergeev_2008}.

The {\it internal mechanical energy} term, $U^{(\alpha)}_{mec}$, appearing in Eq.~\ref{eq2} may be defined by:
\begin{equation}\label{eq2a}
U^{(\alpha)}_{mec} = m^{(\alpha)} \phi^{(\alpha)} (\mathbf{r})
+ m^{(\alpha)} \sum_{\substack{\beta=1 \\ \beta \neq \alpha}}^N \phi^{(\alpha,\beta)}.
\end{equation}
Considering that by definition of thermodynamic temperature, $\partial S^{(\alpha)} / \partial U^{(\alpha)} \equiv1/T^{(\alpha)}$, then it follows
\begin{equation}\label{eq2aa}
\pmb{\nabla}_{r^{(\alpha)}} U^{(\alpha)} = - m^{(\alpha)} \pmb{\nabla} \phi^{(\alpha)} - \frac{\mathbf{p}^{(\alpha)}}{m^{(\alpha)}} \cdot \pmb{\nabla} \mathbf{p}^{(\alpha)} - \pmb{\nabla}_{r^{(\alpha)}} \left( \frac{J^{(\alpha)}2}{2I^{(\alpha)}} \right) - q^{(\alpha)} \pmb{\nabla} V^{(\alpha)} + q^{(\alpha)} \pmb{\nabla} (\mathbf{A}^{\alpha} \cdot \mathbf{v}^{(\alpha)}).
 \end{equation}
Eq.~\ref{eq2aa} contains the particle's self-energy and the particle interaction energy for the gravitational and electromagnetic fields, but it may also include other terms representing different occurring phenomena (energy as a bookkeeping concept), such as terms included in Eq.~\ref{eq2a}.
We may here recall that the entropic flux in space is a kind of generalized force
$X_{\alpha}$ ~\cite{Prigogine:71,Khinchin}, and it can be shown that the following equation holds:
\begin{displaymath}
T \pmb{\nabla}_{r^{(\alpha)}} \mathfrak{S}^{(\alpha)} = - q^{(\alpha)} \pmb{\nabla}_{r^{(\alpha)}} V^{(\alpha)}
+ q^{(\alpha)} \pmb{\nabla}_{r^{(\alpha)}} (\mathbf{A}^{(\alpha)} \cdot \mathbf{v}^{(\alpha)})
\end{displaymath}
\begin{equation}\label{eq2aaa}
+ m^{(\alpha)} \mathbf{v}^{(\alpha)} \cdot \pmb{\nabla} \mathbf{v}^{(\alpha)}
- \pmb{\nabla}_{r^{(\alpha)}} \left( \frac{(\pmb{J}^{(\alpha)})^2}{2 I^{(\alpha)}} - \pmb{\omega} \cdot \pmb{J}^{(\alpha)} \right),
\end{equation}
Now, it make sense to write the fundamental equation of thermodynamics under the form of a spacetime differential equation:
\begin{equation}\label{eq2ab}
T \pmb{\nabla} \overline{S} + \sum_{\alpha} m^{(\alpha)} \frac{\partial \mathbf{v}^{(\alpha)}}{\partial t} = \sum_{\alpha} \pmb{\nabla} \mathbf{U}^{(\alpha)}.
\end{equation}
Taking into account the convective derivative, $d\mathbf{v}^{(\alpha)}/dt \equiv \partial \mathbf{v}^{(\alpha)}/\partial t + \mathbf{v}^{(\alpha)} \cdot \pmb{\nabla} \mathbf{v}^{(\alpha)}$, after some algebra we obtain:
\begin{displaymath}
m^{(\alpha)} \frac{d \mathbf{v}^{(\alpha)}}{d t} = - T \pmb{\nabla}_{r^{(\alpha)}} \mathfrak{S}^{\alpha} - m^{(\alpha)} \pmb{\nabla} \phi^{(\alpha)} - q^{(\alpha)} \pmb{\nabla} V^{(\alpha)}
\end{displaymath}
\begin{equation}\label{eq2b}
+ q^{(\alpha)} \pmb{\nabla} (\mathbf{A}^{(\alpha)} \cdot \mathbf{v}^{(\alpha)})
- \pmb{\nabla}_{r^{(\alpha)}} \left( \frac{(\mathbf{J}^{(\alpha)})^2}{2 I^{(\alpha)}} - \pmb{\omega} \cdot \mathbf{J}^{(\alpha)} \right) + \mathbf{F}^{(\alpha)}_{ext}.
\end{equation}

For conciseness, the term $U^{(\alpha)}$ now includes all forms of energy inserted into the above Eq.~\ref{eq2}. On the right-hand side (r.h.s.), the first term must be present whenever the mechanical and thermodynamical equilibrium conditions are not fulfilled; the second term is the gravitational force term; the third and fourth terms constitute the Lorentz force; the fifth term is a new term which represents the transport of angular momentum; while the last term represents other external forces not explicitly included and actuating over the particle $(\alpha)$.

\subsection{Examples:}

\subsubsection{Extended fundamental equation of dynamics}

The present formalism was applied in a previous work~\cite{Pinheiro:04}, and therein we obtained the ponderomotive forces acting on a charged particle. For a neutral particle (body) in a gravitational field, Eq.~\ref{eq2b} points to a kind of extended fundamental equation of dynamics for a given species $(\alpha)$ at equilibrium and at a given point of space-time (Eulerian description):
\begin{equation}\label{eq3}
m^{(\alpha)} \frac{\partial \mathbf{v}^{(\alpha)}}{\partial t} = -m^{(\alpha)} \pmb{\nabla} \phi^{(\alpha)} - \pmb{\nabla}_{r^{(\alpha)}} \left( \frac{\mathbf{J}^{(\alpha)2}}{2 I^{(\alpha)}} - \pmb{\omega} \cdot \mathbf{J}^{(\alpha)} \right).
\end{equation}
Eq.~\ref{eq3} has a new term because the body possess an intrinsic angular momentum. In a non-rotating frame of reference we must put $\omega=0$, and from the work-energy theorem it is obtained the total mechanical energy of the system, $\mathcal{E}_{mec}=K+U+J_c^2/2I_c$. This is the common approach in classical mechanics. We are interested in the effect of a given force in a given point of space-time, not in its effect along the particle trajectory. It is worth to point out that Eq.~\ref{eq3} was obtained through a variational procedure in contrast to the usual conservation theorem used, for example, in Ref.~\cite{Liboff,Huang}.

\vspace{1.0cm}

Included in the internal energy term are the interpressure term, see Eq.~\ref{eq2a} (we consider here a
homogeneous and isotropic fluid). The above described framework (see also Ref.~\cite{Pinheiro:04} for additional information) leads us to the well-known
hydrodynamic equation for a given species $(\alpha)$:
\begin{displaymath}
m^{(\alpha)} \frac{d \mathbf{v}^{(\alpha)}}{dt} =
- \pmb{\nabla}_{r^{(\alpha)}} (m^{(\alpha)} \phi^{(\alpha)})
- m^{(\alpha)} \pmb{\nabla}_{r^{(\alpha)}} \sum_{\substack{\beta=1\\ \beta \neq \alpha}}^N \phi^{(\alpha,\beta)}
\end{displaymath}
\begin{equation}\label{eq4}
- \pmb{\nabla}_{r^{(\alpha)}} \left( \frac{J^{(\alpha)2}}{2I^{(\alpha)}} - \pmb{\omega} \cdot \mathbf{J}^{(\alpha)} + T \mathfrak{S}^{(\alpha)} \right).
\end{equation}
Here, in the r.h.s. of the Eq.~\ref{eq4} we introduce explicitly external forces terms, eventually present in open systems.

Using the following correspondence from particle to fluid description
\begin{equation}\label{eq5}
\sum_{\alpha} m^{(\alpha)} \rightarrow \int_V d^3 x^{'}
\rho_v(\mathbf{x'}),
\end{equation}
and, as well, a analogue relationship for the electric charge
\begin{equation}\label{eq5a}
\sum_{\alpha} q^{(\alpha)} \rightarrow \int_V d^3 x^{'} \rho(\mathbf{x'}),
\end{equation}
we can can rewrite Eq.~\ref{eq4} under the form of the Euler (governing)
equation:
\begin{equation}\label{eq6}
\rho_v \frac{d \mathbf{v}}{d t} = - \rho_v \pmb{\nabla}_r \phi - \pmb{\nabla}_r p - \pmb{\nabla}_r \Phi_J - \pmb{\nabla}_r f.
\end{equation}
Here, as usual, the total interparticle pressure term (e.g., Ref.~\cite{Chandrasekhar1}) is given by:
\begin{equation}\label{eq7}
p (\mathbf{r})= \sum_{\alpha} m^{(\alpha)} \sum_{\substack{\beta=1\\\beta \neq \alpha}}^N
\phi^{(\alpha,\beta)}(\mathbf{r}).
\end{equation}
To simplify, we introduce a functional integral in the form of an intrinsic angular momentum energy density (comprising the ``interaction energy term", $\pmb{\omega} \cdot \mathbf{J}$), $\Phi_J$:
\begin{equation}\label{eq7a}
\sum_{\alpha} \left[ \frac{J^{(\alpha)2}}{2I^{(\alpha)}} - \pmb{\omega} \cdot \mathbf{J}^{(\alpha)} - (\Delta F)^{(\alpha)} \right] \to \int [\Phi_J(\mathbf{x}') + f(\mathbf{x}')] d^3 x',
\end{equation}
considering that the intrinsic angular momentum density refers to a given blob of fluid (with inertial momentum $I$, a measure of the local rotation, or spin, of the fluid element),and its associated free energy $f=f_0-Ts$ (per unit of volume). Eq.~\ref{eq6} also means that function $S^{(\alpha)}$ (the {\it field integral of $\mathbf{r}^{(\alpha)}$}) is constant along the integrals curves of the space field $\mathbf{r}^{(\alpha)}$. The gradient of the free energy $f$ of the out of equilibrium state is the source of spontaneous change from an unstable state to a more stable state, while performing work. For example, a common source of free energy in a collisionless plasma is an electric current~\cite{Gary_2005}; in a magnetically confined plasma, several classes of free energy sources are available to drive instabilities, e.g., relaxation of a non-Maxwellian, nonisotropic velocity distribution~\cite{Stacey_2005}. At this stage, it is worthwhile to refer that our procedure includes the effect of angular momentum (through Eq.~\ref{eq2}), as it should be in a consistent theory, according to Curtiss~\cite{Curtiss 1956}.

Using, furthermore, the mathematical identity:
\begin{equation}\label{eq10a}
\pmb{\nabla} (\mathbf{A}^{(\alpha)} \cdot \mathbf{v}^{(\alpha)}) = (\mathbf{A}^{(\alpha)} \cdot \pmb{\nabla})\mathbf{v}^{(\alpha)} + (\mathbf{v}^{(\alpha)} \cdot \pmb{\nabla})\mathbf{A}^{(\alpha)} + [\mathbf{A}^{(\alpha)} \times [\pmb{\nabla} \times \mathbf{v}^{(\alpha)}]]+ [\mathbf{v}^{(\alpha)} \times [\pmb{\nabla} \times \mathbf{A}^{(\alpha)}]],
\end{equation}
we obtain after some algebra the following expression:
\begin{equation}\label{eq10b}
\pmb{\nabla} \left( \mathbf{A}^{(\alpha)} \cdot \mathbf{v}^{(\alpha)} \right) = -\frac{\partial \mathbf{A}^{(\alpha)}}{\partial t} - \left[ \pmb{\omega} \times \mathbf{A}^{(\alpha)} \right] + \left[ \mathbf{v}^{(\alpha)} \times \mathbf{B}^{(\alpha)} \right].
\end{equation}
Here, $\mathbf{B}=[\pmb{\nabla} \times \mathbf{A}]$. In addition, we may notice that the following equality holds:
\begin{equation}\label{eq7aaa}
(\mathbf{A}^{(\alpha)} \cdot \pmb{\nabla}) \mathbf{v}^{(\alpha)} = [\pmb{\omega} \times \mathbf{A}^{(\alpha)}],
\end{equation}
where
\begin{equation}\label{eq11a}
\mathbf{A}^{(\alpha)} = \sum_{\substack{\beta=1\\\beta \neq \alpha}} q^{(\beta)} \frac{\mathbf{v}^{(\beta)}}{r_{\alpha \beta}},
\end{equation}
denotes the vector potential actuating on $(\alpha)$ particle due to all the other particles, and
the vorticity is defined by
\begin{equation}\label{eq12}
\pmb{\Omega^{(\alpha)}} = [\pmb{\nabla}_{\mathbf{r}^{(\alpha)}} \times
\mathbf{v}^{(\alpha)}] = 2 \pmb{\omega^{(\alpha)}}.
\end{equation}

Therefore, it follows the general equation of dynamics for a physical system (Lagrangian description):
\begin{equation}\label{eq7b}
\rho \frac{d \mathbf{v}}{d t} = \rho \mathbf{E} + [\mathbf{J} \times \mathbf{B}] - \pmb{\nabla} \phi -
\pmb{\nabla} p + \rho [\mathbf{A}  \times \pmb{\omega}].
\end{equation}
The last term in the r.h.s. of the Eq.~\ref{eq7b} is a new term that represents a kind of topological spin vector~\cite{Kiehn_2007}, an artifact of nonequilibrium process. As it will be shown subsequently, the topological spin vector plays a role in plasma arcs, and as well in magnetocumulative generators~\cite{Onoochin_2003}, and suggests a new method to obtain the helicity transport equation~\cite{Ishida:97}.

\subsubsection{Rigid body rolling down an inclined plane with holonomic constraint}

We may check Eq.~\ref{eq3} in a standard example of classical mechanics: a rigid body of mass $M$ rolling down an inclined plane making an angle $\theta$ with the horizontal (see, e.g., p. 97 of Ref.~\cite{Lamb}). Eq.~\ref{eq4} can be applied to solve the problem, with $\omega=0$ (there is no rotation of the frame of reference) and considering that only the gravitational force acts on the rolling body, with inertial moment relative to its own center of mass given by $I_c=\beta MR^2$. Hence, we obtain:
\begin{equation}\label{eq7c}
M \ddot{x}= Mg \sin \theta - \partial_x w.
\end{equation}
Here, $w \equiv \frac{(\mathbf{J_c})^2}{2 I_c}$. Assuming that the x-axis is directed along the inclined plane, and considering that the angular momentum relative to the rigid body center of mass is given by $J_c=I_c \omega '$, with $\omega' = d\theta /dt$, and noticing that $d x = v_x dt$ (holonomic constraint), it is readily obtained
\begin{equation}\label{eq7d}
M \ddot{x}= M a_x = M g \sin \theta - I_c \omega' \frac{d \omega'}{v_x d t} = Mg \sin \theta - \beta MR^2\frac{\omega'}{\omega' R} \alpha.
\end{equation}
Since $\alpha=a/R$, then it results the well-known equation
\begin{equation}\label{eq7e}
a_x=\frac{g \sin \theta}{(1+\beta)}.
\end{equation}

\section{Equilibrium and stability of a rotating plasma}

Extremum conditions imposed on entropy or internal energy, does not only
provide criteria for the evolution of the system, but, in addition, they determine the stability of thermodynamic systems, at equilibrium. Further, it
has been shown~\cite{Lavende:74} that a state of mechanical
equilibrium can be reached, if the entropy increases with
distance:
\begin{equation}\label{eq9}
T \mathrm{L}_r \overline{S} >0,
\end{equation}
where $\mathrm{L}_r$ is the Lie derivative along the vector field $\mathbf{r}$ acting on scalar $\overline{S}$.
Therefore, we can search another extremal condition through the general expression
\begin{widetext}
\begin{equation}\label{eq10}
T \frac{\partial \mathfrak{S}^{(\alpha)}}{\partial
\mathbf{r}^{(\alpha)}} = - q^{(\alpha)}
\pmb{\nabla}_{\mathbf{r}^{(\alpha)}} V^{(\alpha)} + q^{(\alpha)}
\pmb{\nabla}_{\mathbf{r}^{(\alpha)}} (\mathbf{v}^{(\alpha)} \cdot
\mathbf{A}^{(\alpha)}) - m^{(\alpha)} \frac{\partial
\mathbf{v}^{(\alpha)}}{\partial t}
- \pmb{\nabla}_{r^{(\alpha)}} \left( \frac{(\mathbf{J}^{(\alpha)})^2}{2 I^{(\alpha)}} - \pmb{\omega} \cdot \mathbf{J}^{(\alpha)} \right).
\end{equation}
\end{widetext}

According to Noether's theorem, the total canonical momentum is
conserved in a closed system, and hence we can state the {\it closure
relation}:
\begin{equation}\label{eq13}
\sum_{\alpha=1}^N \left[\mathbf{p}^{(\alpha)} + q^{(\alpha)}
\mathbf{A}^{(\alpha)} \right]=0.
\end{equation}
It can be shown (see Ref.~\cite{Pinheiro:02,Pinheiro:04} for
details) that the relationship that prevails in equilibrium on a
rotating plasma is given by
\begin{widetext}
\begin{equation}\label{eq11}
q^{(\alpha)} \mathbf{E}^{(\alpha)} + q^{(\alpha)} [\mathbf{v} \times
\mathbf{B}^{(\alpha)}] = m^{(\alpha)} \pmb{\nabla}_{\mathbf{r}^{(\alpha)}}
\phi^{(\alpha)} + m^{(\alpha)} \sum_{\substack{\beta=1\\\beta \neq \alpha}}
\pmb{\nabla}_{\mathbf{r}^{(\alpha)}} \phi^{(\alpha,\beta)} - q^{(\alpha)}
[\mathbf{A}^{(\alpha)} \times \pmb{\omega}^{(\alpha)}].
\end{equation}
\end{widetext}

We may develop the correspondence drawn from Eqs.~\ref{eq5}-~\ref{eq5a} and to
obtain the {\it general condition of equilibrium of a rotating
plasma}, in the presence of gravitational and electromagnetic
interactions (e.g., Ref.~\cite{Chakraborty_78}):
\begin{equation}\label{eq15}
\rho \mathbf{E} + [\mathbf{J} \times \mathbf{B}] = \pmb{\nabla} \Phi +
\pmb{\nabla} p - \rho [\mathbf{A}  \times \pmb{\omega}].
\end{equation}
Hereinabove, to simplify the algebra, we took the averaged angular frequency $\pmb{\omega}$, for the all system.
Embedded into Eq.~\ref{eq15}, it is present the vector potential $\mathbf{A}$, on the same footing as the $E$ and $B$-fields. The importance of the vector potential  when compared to the others, depends on the kind of $B$-field prevalent on the system; for instance, if the $B$-field is homogeneous, the vector potential field predominates on the region near the axis, since $\rho \omega A/ JB \sim 1/r^2$. The topological spin vector term is fundamental, since it produces work, responsible for the system angular momentum modification, producing a rocket-like rotation effect on the plasma. The theoretical framework delineated here, may help to clarify problems related to rotating-plasma systems~\cite{Wilcox 1959, Slepian 1960}, and controlled thermonuclear plasma confinement ~\cite{Hurricane 1998}.

\subsection{Bennett pinch}

\begin{figure}
  \includegraphics[width=3 in]{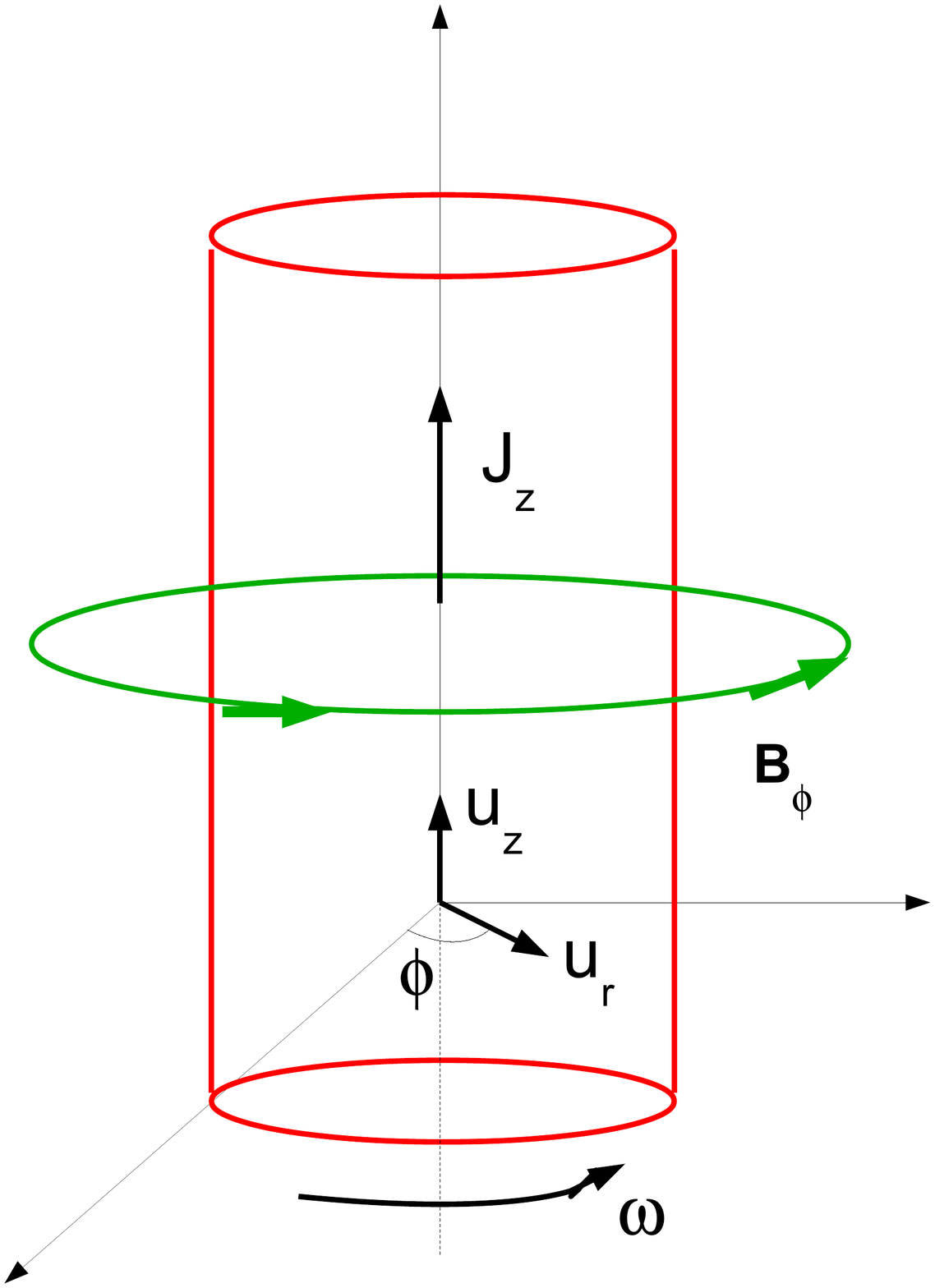}\\
  \caption{Geometry and vectorial fields in the Bennett pinch generated by an axial current $J_z$ creating a toroidal field $B_{\phi}$. If, instead, we consider a vacuum arc discharge with radial current $J_r$ and magnetic field $B_z$, we may have rotating arc with angular velocity $\omega$.}\label{fig1}
\end{figure}

The compression of an electric current by a magnetic field, the z-pinch effect, can be studied on the basis of Eq.~\ref{eq15}, which gives the condition for dynamic equilibria. Let us here assume a typical geometry for an infinitely long axisymmetric cylindrical arc (Fig.~\ref{fig1}), with axial current density $J_z=J_z(r)$. Since the current density is assumed constant, Maxwell's equation in the steady state yields the azimuthal component $B_{\theta}=\mu_0 J_z r/2$, for $r \leq R$, with $R$ the outer boundary of the cylindrical arc. The vector potential is purely radial, given by $A_z(r)=-\mu_0 R^2 J_z/4$, for $r<R$, and the Coriolis term plays no role. We can write Eq.~\ref{eq15} under the form:
\begin{equation}\label{eq15a}
    -J_z B_{\theta} = \frac{dp}{dr}.
\end{equation}
and it follows that
\begin{equation}\label{eq15b}
p(r)=\int_r^R \frac{dp}{dr} dr = \frac{1}{4}\mu_0 J_z^2 (R^2 - r^2),
\end{equation}
which is a well-known result.

The interaction between vacuum arcs and transverse magnetic fields, is used in switching devices (see e.g. Refs.~\cite{Klajn 1999,Flurscheim}). We can consider here instead, a coaxial configuration with a cathode on axis with a stabilizing magnetic induction field $\mathbf{B}$, directed along the symmetry axis, and an arc current density $\mathbf{J}$ flowing radially (and assuming a "filamentary" current with radius $R'$, $A_r=-\mu_0 R'^2 J_r/4$, with $\mu_0$ the permeability of the vacuum). In this case, we may apply Eq.~\ref{eq15} to obtain the pressure differential, from the axis to the wall (at $R$):
\begin{equation}\label{eq15c}
\Delta p(r=R)=2 \pi R \left[ -J_r B_z + \mid \rho_c \mid \pi R^2 \frac{\mu_0}{4 \pi} J_r \omega \right] \Delta \theta.
\end{equation}

\begin{subequations}\label{eq15d}
\begin{align}
    \mid \rho_c \mid > \frac{\mu_0}{4 \pi} \frac{B_z}{S} \omega ,  & \;\;\; \text{retrograde rotation} \\
    \mid \rho_c \mid < \frac{\mu_0}{4 \pi} \frac{B_z}{S} \omega ,  & \;\;\; \text{amperian rotation} .
\end{align}
\end{subequations}


Here, $S$ denotes the filamentary current cross-section, and $\rho=\rho_c$. For negative charge carriers $\rho_c=-\mid \rho_c \mid=-en_e$, and we obtain an amperian (clockwise) rotation for high magnetic fields, and relatively weak arc currents, or instead, a retrograde rotation for higher intensity arcs (higher $S$) and small transverse magnetic fields, in agreement with experimental evidence (e.g., Ref.\cite{Klajn 1999}). We may see in this effect the compromise between two different tendencies, the energy which tends to a minimum and the entropy which tends to a maximum. From Eq.~\ref{eq15c} it is obtained an expression for the spot velocity in a transverse magnetic field:
\begin{equation}\label{eq15dd}
v=\omega R=\frac{4 \pi}{\mu_0}\frac{1}{\mid \rho_c \mid} \left( \frac{\epsilon_0 i}{2 \pi \sigma_c} + \frac{B_z R}{S} \right).
\end{equation}
Here, we have made use of the Bernoulli relation, $\Delta p=\epsilon_0 E^2/2$, and of the constitutive equation $J_r=\sigma_c E$, $\sigma_c$ denoting the plasma electrical conductivity. Although Eq.~\ref{eq15dd} is not self-consistent, it shows that the force term $[\mathbf{J} \times \mathbf{B}]$, is not playing alone its key role, and the spot velocity depends linearly of the arc current (see, e.g., Ref.~\cite{Boxman 1995}). A better understanding of this phenomena is crucial, since arc discharges are powerful generators of non-equilibrium atmospheric pressure plasma.

\section{Flux of angular momentum}

We now address the transport of angular momentum, a phenomena of huge importance in several phenomena, such as in the working mechanism of an accretion disk, the formation of a tornado, or the atmospheric circulation.

The equation of conservation of angular momentum can be obtained
from the following equation:
\begin{widetext}
\begin{displaymath}
m^{(\alpha)} \frac{d \mathbf{v}^{(\alpha)}}{dt} =
- T \frac{\partial \mathfrak{S}^{(\alpha)}}{\partial \mathbf{r}^{(\alpha)}}
- \pmb{\nabla}_{\mathbf{r}^{(\alpha)}} (m^{(\alpha)} \Phi^{(\alpha)})
- q^{(\alpha)} \pmb{\nabla}_{\mathbf{r}^{(\alpha)}} V^{(\alpha)}
\end{displaymath}
\begin{equation}\label{eq16}
+ q^{(\alpha)} \pmb{\nabla}_{\mathbf{r}^{(\alpha)}} (\mathbf{v}^{(\alpha)}
\cdot \mathbf{A}^{(\alpha)})
- \pmb{\nabla}_{\mathbf{r}^{(\alpha)}}
\left( \frac{(\mathbf{J}^{(\alpha)})^2}{2I^{(\alpha)}} - \pmb{\omega}^{\alpha} \cdot
\mathbf{J}^{(\alpha)} \right).
\end{equation}
\end{widetext}
Multiplying the previous Eq.~\ref{eq16} by the particle velocity $v_i^{(\alpha)}$ ($i=x,y,z$), and after rearranging the terms, we obtain:
\begin{displaymath}
\frac{1}{2}\sum_{\alpha} m^{(\alpha)} \frac{d
}{dt}|v_i^{(\alpha)}|^2 = -\frac{1}{2} G \sum_{\alpha} \sum_{\beta
\neq \alpha} m^{(\alpha)} m^{(\beta)}
\frac{(x_i^{(\alpha)}-x_i^{(\beta)})(v_i^{(\alpha)}-v_i^{(\beta)})}{|\mathbf{x}^{(\alpha)}
- \mathbf{x}^{(\beta)}|^3}
\end{displaymath}
\begin{displaymath}
-\frac{1}{2} \sum_{\alpha} \sum_{\beta \neq \alpha} \frac{1}{4 \pi \varepsilon_0}q^{(\alpha)}
q{(\beta)}\frac{(x_i^{(\alpha)}-x_i^{(\beta)})(v_i^{(\alpha)}-v_i^{(\beta)})}{|\mathbf{x}^{(\alpha)}
- \mathbf{x}^{(\beta)}|^3}
- \frac{1}{2}\sum_{\alpha} (q^{(\alpha)} v_i^{(\alpha)} \frac{\partial A_i^{(\alpha)}}{\partial t})
\end{displaymath}
\begin{displaymath}
+ \frac{3}{2}\sum_{\alpha}\varepsilon_{ijk}\omega_k (q^{(\alpha)} A_j^{(\alpha)} v_i^{(\alpha)})
+ \frac{1}{2}\sum_{\alpha}\varepsilon_{ijk} (q^{(\alpha)} v_i^{(\alpha)} v_j^{(\alpha)} B_k^{(\alpha)})
\end{displaymath}
\begin{equation}\label{eq18}
+ \frac{1}{2}\sum_{\alpha}v_i^{(\alpha)} \frac{\partial }{\partial x_i^{(\alpha)}} \left[ \frac{\mid \mathbf{J^{(\alpha)}} \mid^2}{2 I^{(\alpha)}}
- (\pmb{\omega}^{\alpha} \cdot \mathbf{J}^{(\alpha)}) \right]
+ T \sum_{\alpha} v_i^{(\alpha)} \frac{\partial \mathfrak{S}^{(\alpha)}}{\partial x_i^{(\alpha)}}.
\end{equation}
Eq.~\ref{eq18} can be written in a more comprehensive form, if we
separate the terms from different contributions. For this purpose, let's define the
total kinetic energy by means of the expression
\begin{equation}\label{eq19}
\mathfrak{I}\equiv \frac{1}{2}\sum_{\alpha} m^{(\alpha)}
|\mathbf{v}^{(\alpha)}|^2,
\end{equation}
and let us denote by
\begin{equation}\label{eq20}
\mathfrak{P}_g \equiv \frac{1}{2}\sum_{\alpha}
m^{(\alpha)}\varphi^{(\alpha)}=\frac{1}{2}G\sum_{\alpha} \sum_{\beta
\neq \alpha} \frac{m^{(\alpha)} m^{(\beta)}}{|\mathbf{x}^{(\alpha)}
- \mathbf{x}^{(\beta)}|},
\end{equation}
the overall gravitational energy. Similarly, the total electrostatic
energy is given by:
\begin{equation}\label{eq21}
\mathfrak{P}_v \equiv \frac{1}{2}\frac{1}{4 \pi \varepsilon_0}
\sum_{\alpha}\sum_{\beta \neq \alpha}\frac{q^{(\alpha)}
q^{(\beta)}}{|\mathbf{x}^{(\alpha)} - \mathbf{x}^{(\beta)}|}.
\end{equation}
Following now the Umov-Poynting procedure, it can be shown that the previous Eq.~\ref{eq16} reduces after some algebra to the following one:
\begin{displaymath}
\frac{d}{dt}(\mathfrak{I}) = - \frac{d}{dt}(\mathfrak{P}_g)-\frac{d}{dt}\mathfrak{P}_v
+ \frac{3}{2}\sum_{\alpha}q^{\alpha}[\mathbf{v}^{(\alpha)} \times [\mathbf{A}^{(\alpha)} \times \pmb{\omega}]]
\end{displaymath}
\begin{equation}\label{eq22}
+ \frac{1}{2}\sum_{\alpha}\mathbf{v}^{(\alpha)} \cdot \pmb{\nabla}_{\mathbf{r}^{(\alpha)}}
\left[\frac{|\mathbf{J}^{\alpha}|^2}{2I^{(\alpha)}}
- (\pmb{\omega} \cdot \mathbf{J}^{(\alpha)}) \right]
+ T \sum_{\alpha} v_i^{(\alpha)} \frac{\partial \mathfrak{S}^{(\alpha)}}{\partial x_i^{(\alpha)}}.
\end{equation}

As we did before, it is convenient to introduce a new physical quantity, representing the rotational energy:
\begin{equation}\label{eq23}
\mathfrak{P}_{rot}=-\frac{3}{2} \sum_{\alpha} q^{\alpha}
[\mathbf{v}^{(\alpha)} \times [\mathbf{A}^{(\alpha)} \times
\pmb{\omega}]].
\end{equation}
Then, the above Eq.~\ref{eq22}, can be written under the more general form:
\begin{equation}\label{eq24}
\frac{d }{d t}(\mathfrak{I} + \mathfrak{P}_g +
\mathfrak{P}_v + \mathfrak{P}_{rot}) = \frac{1}{2} \mathbf{v} \cdot \pmb{\nabla}_{\mathbf{r}} \left[
\frac{|\mathbf{J}|^2}{2 I} - (\pmb{\omega} \cdot \mathbf{J}) - \Delta F \right].
\end{equation}
Here, $\mathbf{v}$ represents now the velocity of an ``element of fluid", and the last term in brackets is an average local value. A simple analysis of the Eq.\ref{eq24} shows that the system is
stable in rotatory motion, provided that the following condition is satisfied:
\begin{equation}\label{eq25}
\pmb{\nabla}_r \left[
\frac{|\mathbf{J}|^2}{2I}- (\pmb{\omega}
\cdot \mathbf{J}) - \Delta F \right] < 0.
\end{equation}
The last term is the free energy per unit of volume, $F=F_0- T S$. Remark that Eq.~\ref{eq25} is consistent with Gibbs distribution in a rotating body (see, e.g., Ref. ~\cite{Landau_1}), which means that the radial flux of energy must be positive, flowing out radially from the system's boundary. In addition, we notice that the equilibrium of a gravito-electromagnetic system depends upon its mechanical rotational properties, but also on the free energy available, linking intrinsically any mechanical process to thermodynamic variables, and opening options for possible unconventional mechanisms to control instabilities. In the domain of astrophysical plasmas, gravity and rotation usually dominate over the magnetic field of force, being crucial for the development of instabilities.

From Eq.~\ref{eq24} we may see that equilibrium ensues (neglecting thermal and configurational effects) when the velocity of rotation of the fluid satisfies the local condition (e.g., Ref.~\cite{Balbus 1991}):
\begin{equation}\label{eq26a}
\frac{d \Omega(r)^2}{dr} \geq 0,
\end{equation}
where we identified $\omega$ with the bulk angular velocity. Eq.~\ref{eq26a} is related to the conservation of energy.
However, when condition~\ref{eq25} is not fulfilled, it gives rise to the magnetorotational instability (MIR)~\cite{Balbus 1991,Velikhov 2005}, that appears as the result of the interplay of its three different terms: i) the angular momentum acquired by the fluid (or particles); ii) an interaction term due to the coupling between the fluid angular momentum with the driven angular velocity; iii) the fluid thermal energy and configurational entropy.

A typical experiment consists in the rotation of a fluid between two concentric cylinders - related to the so called Taylor-Couette instability - driven by velocity gradients. In the presence of an axial magnetic field, the Taylor-Couette instability is onset when Eq.~\ref{eq26a} is not satisfied. Also, we may expect that, owing to the fact that for two different species with different inertial moment, $I^{\alpha} \neq I^{\beta}$, it can be expected that at some point $r=r_c$ of the radial axis an inversion of the sign of the inequality of Eq.~\ref{eq25} must take place, and instability is onset. In particular, in the presence of two different species with different inertial moments, the fluid may well be intrinsically unstable at enough high angular speed. MIR instabilities concur against the stability of plasma configurations and in the 1960's, when MHD power plants were considered to be an economic process to convert thermal energy into electrical energy, E. Velikhov, in 1962, discovered the electrothermal instability which is at the source of strong magnetohydrodynamic turbulence~\cite{Velikhov_1959,Chandrasekhar_1960,Balbus 1991}.

The above Eq.~\ref{eq24} can be written in the form of a conservative equation
for energy:
\begin{displaymath}
\frac{\partial }{\partial t}(\mathfrak{I}+ \mathfrak{P}_g +
\mathfrak{P}_v + \mathfrak{P}_{rot}) = -\frac{1}{2} \sum_{\alpha}
\pmb{\nabla}_{\mathbf{r}^{(\alpha)}}  \cdot \left\{ \mathbf{v}^{(\alpha)}
\left[ -\frac{|\mathbf{J}^{(\alpha)}|^2}{2 I^{(\alpha)}}
+ (\pmb{\omega} \cdot \mathbf{J}^{(\alpha)})
- T \mathfrak{S}^{(\alpha)} \right] \right\}
\end{displaymath}
\begin{equation}\label{eq27}
+ \frac{1}{2} \sum_{\alpha} \left[ -\frac{|\mathbf{J}^{(\alpha)}|^2}{2
I^{(\alpha)}} + (\pmb{\omega} \cdot \mathbf{J}^{(\alpha)})
- T \mathfrak{S}^{(\alpha)} \right] (\pmb{\nabla}_{\mathbf{r}^{(\alpha)}} \cdot \mathbf{v}^{(\alpha)}).
\end{equation}
Finally, we can transform the above Eq.~\ref{eq27} into a {\it
Poynting's Theorem for rotating fluids}:
\begin{equation}\label{eq28}
-\frac{\partial }{\partial t}(\mathcal{U}) = \pmb{\nabla} \cdot \mathbf{S} +
\mathcal{P}',
\end{equation}
after defining a kind of {\it Poynting vector for rotational fluids}, which gives the rate of rotational energy flow:
\begin{equation}\label{eq29}
\mathbf{S}=\frac{1}{2} \sum_{\alpha} \mathbf{v}^{(\alpha)}\left[
-\frac{|\mathbf{J}^{(\alpha)}|^2}{2 I^{(\alpha)}} +
(\pmb{\omega}^{(\alpha)} \cdot \mathbf{J}^{(\alpha)}) -
T \mathfrak{S}^{(\alpha)} \right].
\end{equation}
The {\it power input} driving the process (source$/$sink term) is given by:
\begin{equation}\label{eq30}
\mathcal{P}' \equiv \frac{1}{2} \sum_{\alpha} (\pmb{\nabla} \cdot \mathbf{v}^{(\alpha)}) \left[
\frac{|\mathbf{J}^{(\alpha)}|^2}{2 I^{(\alpha)}} - (\pmb{\omega}
\cdot \mathbf{J}^{(\alpha)})
+ T \mathfrak{S}^{(\alpha)} \right],
\end{equation}
and the total energy is defined by summing up the different contributions:
\begin{equation}\label{eq30a}
\mathcal{U} = \mathfrak{I} + \mathfrak{P}_g + \mathfrak{P}_v + \mathfrak{P}_{rot}.
\end{equation}

Here, the term $T \mathfrak{S}^{(\alpha)}$ represents the thermal energy associated to specie $(\alpha)$,
equal to $-\Delta F$, the free energy of the physical system. For
a system in contact with a reservoir at constant temperature this is
the maximum work which can be done, the free energy
tends to decrease for a system in thermal contact with a heat
reservoir. In particular, notice that when
the angular velocity $\omega$ is multiplied by Eq.~\ref{eq16}, it is obtained the driving power. It is worth to point out the presence of the term $\pmb{\omega} \cdot \mathbf{J}$ which plays an analogue role to the slip in electrical induction motors, that is, the lag between the rotor speed and the magnetic field's speed, provided by the stator's rotating speed. Furthermore, we may notice that the power input depends on the fluid compressibility $\pmb{\nabla} \cdot \mathbf{v}$. This means that compressibility is a factor that determines the amount of transported angular momentum through the stress-tensor $\tau_{ij}$, and may be responsible of new driving mechanism in addition to the well-known MRI. The driving energy of the rotating system can be expressed in the form:
\begin{equation}\label{eq30b}
\mathcal{E}_{driv} = \frac{J^2}{2I_p} - (\pmb{\omega} \cdot \mathbf{J}) -\Delta F.
\end{equation}

Next, we will discuss several examples, illustrating the application of the variational method.

\subsection{Power extraction from an hurricane}

A hurricane is a natural airborne structure that converts all forms of kinetic and potential energy by means of the transport of angular momentum from the inner core to the outer regions, conveyed either directly by moving matter, or by non-material stresses such as those exerted by electric fields or magnetic fields~\cite{Valdis}. We may apply Eq.~\ref{eq30} to this specific problem and obtain:
\begin{equation}\label{eq31}
\mathcal{P}'=\omega \left( \frac{J^2}{2I_p} - \pmb{\omega} \cdot \mathbf{J} - \Delta F \right) = -\frac{\partial U_e}{\partial t}
\end{equation}
or
\begin{equation}\label{eq31a}
\mathcal{P}' = \omega \mathcal{E}_{driv}.
\end{equation}
Let us consider specifically the case of a hurricane in an axisymmetric configuration, with $J=\omega I$. Likely we have $\omega^2 I^2/I_p \gg \omega^2 I + \Delta F$. We can envisage a simple and rough model of a hurricane, with total mass $M$ and radius $R$, turning like a solid cylinder, and $I=MR^2/2$. Hence, the total power driving the hurricane is given by
\begin{equation}\label{eq31b}
\mathcal{P}' = \frac{1}{8} \frac{\omega^3}{2} \frac{M^2R^4}{I_p} \propto \omega^3 \frac{M^2R^4}{I_p},
\end{equation}
or, as a function of the fluid density $\rho$, in the form:
\begin{equation}\label{eq31c}
\mathcal{P}' = \frac{\pi^2}{4} \omega^3 \frac{R^6 L^2}{I_p} \rho^2.
\end{equation}
Our result shows the same kind of dependency as shown by Chow $\&$ Chey~\cite{Chow}, and, in particular, it shows that the intrinsic inertial momentum of the particles constituting the fluid plays a substantial role.

\subsection{Radiactive heating of planetary atmosphere and net angular momentum}

It is experimentally shown~\cite{Schubert_69} that periodic radiative heating of the Earth's atmosphere transmits angular momentum to it due to the Earth-atmosphere coupling through frictional interactions ~\cite{AChandrasekar}. The images sent by ESA's Venus Express confirms this fact on Venus (Earth's twin planet), with the presence of a ``double-eye" atmospheric vortex at the planet's south pole and high velocity winds whirling westwards around the planet completing a full rotation in four days.

Schubert and Withehead's~\cite{Schubert_69} did an experiment with the purpose to provide an explanation for the high velocities of apparent cloud formation, in the upper atmosphere of Venus. In this experiment, a bunsen flame rotating below a cylindrical annulus, filled with liquid mercury, induced the rotation of the liquid mercury, in a counter-direction to that of the rotating flame. The speed of the flame was 1 mm$/$s, and the temperature of the mercury increased from room temperature, at the rate of about 3$^0$C per minute. After 5 minutes, a steady-state flow was established, with the mercury rotating in the counter-direction of the flame, with a speed about 4 mm$/$s. Considering that the entire mercury liquid as a spinning body, we can estimate from Eq.~\ref{eq26a} that
\begin{equation}\label{eq31d}
\frac{d}{dr} \left( \frac{I_p \omega'^2}{2} - \frac{I_p \omega^2}{2} + T \mathfrak{S} \right) > 0 \Rightarrow I \Delta(\omega' + \omega)(\omega' - \omega)+\Delta T \mathfrak{S}>0.
\end{equation}
Hence, the following result is obtained:
\begin{equation}\label{eq32}
\omega' \Delta \omega \gtrsim -\frac{1}{2}\frac{\Delta T \mathfrak{S}}{I_p}.
\end{equation}
Here, $\omega$ is the angular speed imposed on the system; $\omega'$ the induced angular speed, due to a sustained source of energy; $\Delta \omega \equiv \omega' - \omega$. We use the following tabulated data: $\mathfrak{S}=16.6$ J.K$^{-1}$mol$^{-1}$ for mercury at temperature of the experiment, $\rho \approx 13.6 \times 10^{-3}$ kg$/$m$^3$, and consider that the volume of mercury is contained in the channel of the experimental apparatus, forming a rim with an average radius $R=30$ cm. With Eq.~\ref{eq26a} we estimate that, after one minute, the speed must be of the order 3.7 mm$/$s. It is clear that the sense of rotation, and its speed, depend on the latent heat stored in the planetary atmosphere and its temperature (through $\Delta T$). Although the initial assumptions taken in the present formulation require further research, considering that interactive terms with the medium, conveyed for example through thermal diffusion coefficient, are not covered by Eq.~\ref{eq26a}, the agreement is reasonable, and offers a possible explanation for this effect. Once again, the problem of radiative atmosphere heating reveals the interesting interplay between energy and entropy, each one attempting a different equilibrium condition.

\section{Conclusion}

The information-theoretic method proposed in this paper constitute an alternate approach applying the concept of maximizing entropy to the problem of out of equilibrium physical systems. It bears some resemblance with Hamiltonian formulation of dynamics, which expresses first-order constraints of the Hamiltonian $H$ in a $2n$ dimensional phase space, revealing the same formal symplectic structure shared by classical mechanics and thermodynamics.
Although the simplifying assumption of isothermal system rules out its ability to explain accurately such problems as, for example, the coherent transport of angular momentum in astrophysics, or a certain kind of laboratory devices (e.g., the Ranque-Hilsch effect), the present method open to understanding certain trends, in particular, predicting theoretically for electromagnetic-gravitational vortex the angular momentum forced transport outwardly from a symmetry axis of rotation. The kind of Umov-Poynting theorem obtained is revealing of the interplay between entropy and energy, the energy tending to a minimum, and the entropy tending to a maximum, explaining the formation of physical structures. In particular, it was shown that compressibility is an important property in the transport of angular momentum, and a possible driving mechanism for instability. This development is seen to be advantageous and opens options for systematic improvements.

\begin{acknowledgments}
We are grateful to Professor Roman W. Jackiw, MIT Center for Theoretical Physics, for keeping us informed of the fundamental paper referred to in [13].
The author gratefully acknowledge partial financial support by the International Space Science Institute (ISSI) as visiting scientist, and express special thanks to Professor Roger Maurice-Bonnet. 
\end{acknowledgments}

\bibliographystyle{amsplain}

\bibliographystyle{apsrev}
\end{document}